\documentstyle[leqno,12pt,graphics,epic,eepic]{article}
\input{psfig.sty}
\let\myendproof\endproof
\def\endproof{{\vbox{\hrule\hbox{%
   \vrule height1.3ex\hskip0.8ex\vrule}\hrule
  }}\par\myendproof}
\def\myendofproof{\vbox{\hrule\hbox{%
   \vrule height1.3ex\hskip0.8ex\vrule}\hrule}\par}

\newtheorem{mytheorem}{Theorem}
\newtheorem{theorem}{Theorem}[section]
\newtheorem{lemma}[theorem]{Lemma}

\newtheorem{definition}{Definition}
\newenvironment{PROOF}{\par\noindent {\em Proof. }}{\myendofproof}

\newenvironment{tabAlgorithm}[2]{
\setcounter{algorithmLine}{1}
\samepage
\begin{tabbing}
999\=\kill
#1 \\
\parbox{4.8in}{\small\sl #2}
}{
\end{tabbing}
}
\newcounter{algorithmLine}
\newcommand{\algline}{\\\thealgorithmLine\hfil\>\stepcounter{algorithmLine}}

\newcommand{\fig}[3] 
{
 \begin{figure}[htbp]
 \begin{center}
 \input{#1.tex}
 \end{center}
 \caption{#3}
 \label{#2}
 \end{figure}
}

\newcommand{\SC}[1]{\mbox{\sc #1}}

\newcommand{\minbeta}[0]{1+\frac{2}{\alpha-1}}

\newcommand{\ABL}[0]{$\bigl(\alpha,\beta\bigr)$-LAST}
\newcommand{\OTL}[0]{$\bigl(\alpha,\minbeta\bigr)$-LAST}

\newcommand{\dist}[3]{\mbox{\it D\/}_{#1}(#2,#3)}
\newcommand{\parent}[2]{\mbox{parent}_{#1}(#2)}

\newcommand{\algboxB}[3]{
\begin{center}
\fbox{\hspace{#2}\parbox{#1}{\vspace{#2}#3
\parbox{#1}{\vspace{#2}}}\hspace{#2}}%
\end{center}}

\newcommand{\FindLast}{
\begin{figure}[htb]

\algboxB{5in}{0.1in}{

\begin{tabAlgorithm}{$\SC{Find-Last}(T_M, T_S, r, \alpha)$}%
{Input: Min.~spanning tree $T_M$, shortest-path tree $T_S$,  vertex $r$,
 $\alpha > 1$.\\
Output: an \OTL\ rooted at $r$.}
\algline $\SC{Initialize}()$
\algline $\SC{DFS}(r)$
\algline {\bf return} tree $T=\{(v,p[v]) \mid v \in V-\{r\}\}$
\end{tabAlgorithm}

\begin{tabAlgorithm}{$\SC{DFS}(u)$}%
{Traverse the subtree of $T_M$ rooted at $u$,
relaxing edges as they are traversed, 
and adding paths from $T_S$ as needed.}
\algline {\bf if} \= $d[u] > \alpha\,\dist{T_S}{r}{u}$
\algline        \>  {\bf then} $\SC{Add-Path}(u)$
\algline {\bf for} \= each child $v$ of $u$ in $T_M$ 
\algline        \> {\bf do} \= $\SC{Relax}(u, v)$
\algline        \>      \> $\SC{DFS}(v)$
\algline        \>      \> $\SC{Relax}(v, u)$
\end{tabAlgorithm}

\begin{tabAlgorithm}{$\SC{Add-Path}(v)$}%
{Relax edges along path from $r$ to $v$ in $T_S$.}
\algline {\bf if} \= $d[v] > \dist{T_S}{r}{v}$ 
\algline        \> {\bf then} \= $\SC{Add-Path}(\parent{T_S}{v})$
\algline        \>      \> $\SC{Relax}(\parent{T_S}{v},v)$
\end{tabAlgorithm}

\vglue-\baselineskip
}
\caption{Algorithm to compute a LAST.}
\label{spt_alg}
\end{figure}}

\newcommand{\Initialize}{
\begin{tabAlgorithm}{$\SC{Initialize}()$}%
{Initialize distance estimates, parent pointers.}
\algline {\bf for} \= each non-root vertex $v$ do
        $p[v] \leftarrow {\bf nil}$; 
        $d[v] \leftarrow \infty$
\algline $d[r] \leftarrow 0$
\end{tabAlgorithm}}

\newcommand{\Relax}{
\begin{tabAlgorithm}{$\SC{Relax}(u,v)$}%
{Check for shorter path to $v$ through $(u,v)$.}
\algline {\bf if} \= $d[v] > d[u] + w(u,v)$
\algline        \>  {\bf then} \= $d[v]\leftarrow d[u] + w(u,v)$
\algline        \>              \> $p[v]\leftarrow u$
\end{tabAlgorithm}}

\title{Balancing Minimum Spanning Trees \\ and Shortest-Path Trees}

\author{Samir Khuller
\thanks{Computer Science Department, University of
Maryland, College Park, MD~20742. E-mail~: {\tt samir@cs.umd.edu}.
Part of this work was done while this author was at University of Maryland
Institute for Advanced Computer Studies (UMIACS)
and was supported by NSF grants CCR-8906949, CCR-9103135 and CCR-9111348.}
\\ University of Maryland
\and Balaji Raghavachari
\thanks{Computer Science Department, Pennsylvania State University,
University Park, PA~16802. E-mail : {\tt rbk@cs.psu.edu}.}
\\ Penn State 
\and Neal Young
\thanks{University of Maryland Institute for Advanced Computer Studies, 
College Park, MD~20742. E-mail : {\tt young@umiacs.umd.edu}.
Supported by NSF grants CCR-8906949 and CCR-9111348.}
\\ University of Maryland
}

\date{ }

\begin{document}

\begin{titlepage}
\maketitle

\begin{abstract}
We give a simple algorithm 
to find a spanning tree
that simultaneously approximates
a shortest-path tree and a minimum spanning tree.
The algorithm provides a continuous trade-off:
given the two trees and a $\gamma > 0$,
the algorithm returns a spanning tree
in which the distance between any vertex and the root of the shortest-path tree
is at most $1+\sqrt{2}\gamma$ times the shortest-path distance,
and yet the total weight of the tree is at most $1+\sqrt{2}/\gamma$
times the weight of a minimum spanning tree.

Our algorithm runs in linear time and
obtains the best-possible trade-off. It
can be implemented on a CREW PRAM
to run in logarithmic time using one processor per vertex.
\end{abstract}


\thispagestyle{empty}
\end{titlepage}

\section{Introduction}
A {\em minimum spanning tree} of an edge-weighted graph
is a spanning tree of the graph of minimum total edge weight.
A {\em shortest-path tree} rooted at a vertex $r$
is a spanning tree such that for any vertex $v$,
the distance between $r$ and $v$ is the same as in the graph.

Minimum spanning trees and shortest-path trees
are fundamental structures in the study of graph algorithms
\cite{CLR,Dijkstra,Kruskal,Prim};
fast algorithms for finding each are known \cite{FT,GGST}.
Typically, the edge-weighted graph $G$ represents a feasible network.
Each vertex represents a site.
The goal is to install links between pairs of sites
so that signals can be routed in the resulting network.
Each edge of $G$ represents a link that can be installed.
The cost of the edge reflects both the cost to install the link
and the cost (e.g., time) for a signal to traverse the link
once the link is installed.
A minimum spanning tree represents the least costly set of links
to install so that all sites are directly or indirectly connected,
while a shortest-path tree represents the set of links
to install so that for each site, 
the cost for a signal to be sent between the site and the root of the tree
is as small as possible.

The goal of a minimum spanning tree is minimum weight,
whereas the goal of a shortest-path tree 
is to preserve distances from the root.
We show that a single tree can approximately achieve both goals.
That is, the cost to install a set of links so that every site
has a short path to the root 
is only slightly more than the cost just to connect all sites.

In the graph in Figure \ref{leadEx},
the weight of the shortest-path tree
is much more than the weight of a minimum spanning tree.
Conversely, in the minimum spanning tree,
the distance between the root and one of the vertices
is much larger than the corresponding shortest-path distance.
Nonetheless, there is a tree which nearly preserves distances from the root
and yet weighs only a little more than the minimum spanning tree.
We call such a tree a {\em Light Approximate Shortest-path Tree} (LAST).
The main result of this paper is that such trees exist in all graphs
and can be found efficiently.

\fig{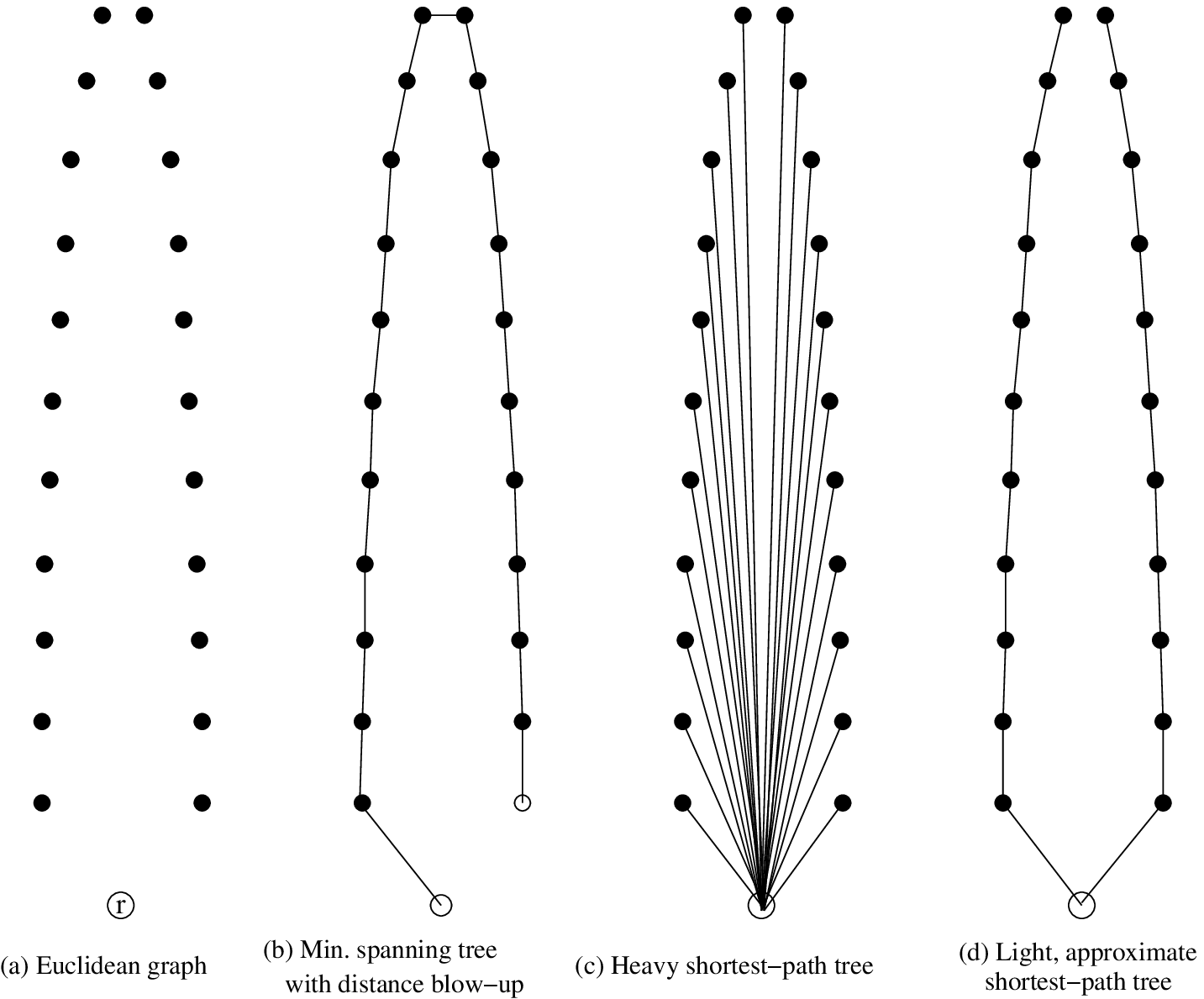}{leadEx}{Approximating
both a minimum spanning tree 
and a shortest-path tree.}

Let $G = (V, E)$ be a graph with non-negative edge weights
and a {\em root} vertex $r$.
Let $G$ have $n$ vertices and $m$ edges.
Let $w(e)$ be the weight of edge $e\in E$.
The {\em distance} $\dist{G}{u}{v}$ 
between vertices $u$ and $v$ in $G$
is the minimum weight of any path in $G$ between them.

\begin{definition}
For $\alpha \ge 1$ and $\beta \ge 1$,
a spanning tree $T$ of $G$ meeting the following two requirements
is called an {\em \ABL}\ rooted at $r$.
\begin{itemize}
\item (Distance)
For every vertex $v$, the distance between $r$ and $v$ in $T$ 
is at most $\alpha$ times the shortest distance from $r$ to $v$ in $G$.
\item (Weight)
The weight of $T$ is at most $\beta$ times the weight of a
minimum spanning tree of $G$.
\end{itemize}
\end{definition}

\begin{mytheorem}[Section \ref{algorithmSec}]
\label{mainthm}
Let $G$ be a graph with non-negative edge weights;
let $r$ be a vertex of $G$; 
let $\alpha > 1$ and $\beta \ge \minbeta$.
Then $G$ contains an {\ABL} rooted at $r$.
The LAST can be computed in linear time
given a minimum spanning tree and a shortest-path tree,
and in $O(m + n\log n)$ time otherwise.
\end{mytheorem}

Note that there is a trade-off between the approximations of the two trees.
The trade-off  is the best possible:
\begin{mytheorem}[Section \ref{optimalitySec}]
\label{optimalityThm}
Fix $\alpha > 1$ and $1 \le \beta < \minbeta$.
There exists a planar graph $G$ with a vertex $r$
such that $G$ contains no {\ABL} rooted at $r$.
Deciding whether a given graph
contains an {\ABL} rooted at a given vertex is NP-complete.
\end{mytheorem}

Note that for $\beta = 1$,
the problem is to find a minimum spanning tree
that best approximates a shortest-path tree.
It follows from Theorem \ref{optimalityThm} that this is NP-complete.
When $\alpha = 1$,
the problem is to find a minimum-weight shortest-path tree.
This problem can be solved in linear time, even in directed graphs:
\begin{mytheorem}[Section \ref{minWtSPTSec}]
\label{minWtSPTThm}
Given any shortest-path tree of a directed or undirected graph
rooted at a given vertex, a
minimum-weight shortest-path tree can be found in linear time.
\end{mytheorem}

Finally, LAST's can also be found quickly in parallel,
given a minimum spanning tree and shortest-path tree
(or approximations thereof, see Section \ref{analysisSec}):
\begin{mytheorem}[Section \ref{parallelSec}]
\label{parallelThm}
Given $\alpha > 1$, a minimum spanning tree, and a shortest-path tree,
an {\OTL}\ can be found
by $n$ processors in $O(\log n)$ time on a CREW PRAM.
\end{mytheorem}

\section{Related Work} 
Trees realizing tradeoffs between weight and distance requirements
were first studied by Bharath-Kumar and Jaffe \cite{BJ}.
The authors' weight requirement was the same as ours,
but their distance requirement 
was that the {\em sum} of the distances from the root to each vertex 
should be at most $\beta$ times the minimum possible sum.
They showed the weaker tradeoff
that the desired tree exists if $\alpha \beta \ge \Theta(n)$.

Awerbuch, Baratz and Peleg \cite{ABP1},
motivated by applications in broadcast-network design,
made a fundamental contribution
by showing that every graph has a {\em shallow-light} tree
--- a tree of {\em diameter} at most a constant times
the diameter of $G$ and of weight at most a constant times
the weight of the minimum spanning tree.
Our algorithm is modification of their algorithm.
Cong, Kahng, Robins, Sarrafzadeh and Wong \cite{CKRSW1,CKRSW2,CKRSW3},
motivated by applications in VLSI-circuit design,
improve the constants in the construction of \cite{ABP1}
and consider variations bounding the {\em radius} of the tree
instead of the diameter.
Recently and independently,
Awerbuch, Baratz and Peleg \cite{ABP2},
modified the algorithm from \cite{ABP1}.
They obtained the same algorithm as in \cite{CKRSW2}
but a stronger analysis, proving that the algorithm computes
an $(\alpha, 1 + \frac{4}{\alpha-1})$-LAST.
Their algorithm takes $O(m +n \log n)$ time.
Our algorithm achieves a strictly stronger distance
requirement than the above algorithms.

Considerable research has been done on 
finding {\em spanners} of small size and weight 
in arbitrary graphs \cite{ADDJ,CDNS,PU}
and in Euclidean graphs (graphs induced by points in the plane)
\cite{CDNS,Chew,Levco,Vaid}.
A $t$-spanner is a low-weight subgraph $G'$ of $G$ 
such that for any two vertices,
the distance between them in $G'$
is at most $t$ times the distance in $G$.
It is known that there are graphs that {\em do not} have constant-spanners
of net weight bounded by a constant times 
the weight of the minimum spanning tree.
Awerbuch, Baratz and Peleg \cite{ABP2} also consider 
light trees that have low {\em average} distance-blowup on all non-tree edges.
References to most of the work on graph spanners may be 
found in the paper by Chandra, Das, Narasimhan and Soares \cite{CDNS}.

We can reduce the problem of finding an {\OTL}
to the problem of finding an $\alpha$-spanner
of weight at most $\left(\minbeta\right)$
times the minimum spanning-tree weight 
in a planar graph.
An algorithm achieving the latter is given in \cite{ADDJ}.
This gives an alternate (but less efficient) method of for finding an {\OTL}.

\section{The Algorithm}
\label{algorithmSec}
The algorithm is given an $\alpha > 1$, a minimum spanning tree,
and a shortest-path tree rooted at a vertex $r$.
It returns an \OTL\ rooted at $r$.

The basic idea of the algorithm 
is to traverse the minimum spanning tree,
maintaining a {\em current tree},
and checking each vertex when it is encountered
to ensure that the distance requirement for that vertex is met
in the current tree.
If it is not met, 
the edges of the shortest path between the vertex and the root 
are added into the current tree.
Other edges are discarded
so that a tree structure is maintained.

After all vertices have been checked
and paths added as necessary, 
the remaining tree is the desired LAST.
The final tree is not too heavy
because a shortest-path is only added 
if the path that it replaces 
is heavier by a factor of $\alpha > 1$.
This allows a charging argument
bounding the net weight of the added paths.

\subsection{Relaxation.  }
The tree is maintained 
by keeping a parent pointer $p[v]$ for each non-root vertex $v$.
To avoid recomputing shortest-path distances when a path is added, 
the algorithm maintains a {\em distance estimate} $d[v]$ for each vertex $v$.
This distance estimate, 
which is an upper bound on the true distance in the current tree,
is used in deciding whether to add a path to the vertex.
The parent pointers and distance estimates are initialized 
and maintained as in \cite[Section 25.1]{CLR}:
\Initialize
\Relax
After executing \SC{Initialize},
the algorithm builds and updates the tree
and maintains the distance estimates
by a sequence of calls to \SC{Relax}. The important invariant maintained
by \SC{Relax} is that 
the edges $\{(p[v],v) : d[v] \ne \infty\}$ form a tree,
with $d[v]$ an upper bound on the distance between the root and $v$ 
in the tree.

\subsection{The algorithm as a sequence of relaxations.  }
\label{relaxSec}
The algorithm performs a depth-first search 
of the minimum spanning tree starting at the root.
For a tree, a depth-first search is simply
an edge-by-edge walk from the root vertex
through the vertices of the tree.
Each edge is traversed twice: once in each direction.
At any time in the search,
the sequence of edges traversed so far forms a walk (a non-simple path)
through the visited vertices.
The walk starts at the root and ends at a vertex
that we call the {\em current vertex}.
We also say the algorithm is {\em visiting} this vertex.

The relaxations done by the algorithm are of two kinds.
The first kind adds shortest paths.
The first time vertex $v$ is visited,
if $d[v]$ exceeds $\alpha$ times the distance from the root to $v$ 
in the shortest-path tree,
then the edges of the shortest path are relaxed as needed
to lower $d[v]$ to the shortest-path distance.

The second kind extends or modifies the current tree
to use a minimum-spanning-tree edge if it is useful.
Specifically, when an edge $(u,v)$ is traversed from $u$ to $v$,
$\SC{Relax}(u,v)$ is called.
This guarantees inductively that $d[v]$
is bounded by the weight of the shortest path from the root $r$ to vertex $v'$
plus the weight of the minimum-spanning-tree path from $v'$ to $v$,
where $v'$ is the most recent vertex 
to have its shortest path added.
This invariant is what allows the weight of the added paths to be bounded.

When the depth-first search finishes,
the current tree is the desired LAST.
The full algorithm is given in Figure \ref{spt_alg}.

\FindLast

\subsection{A sample execution.  }

\fig{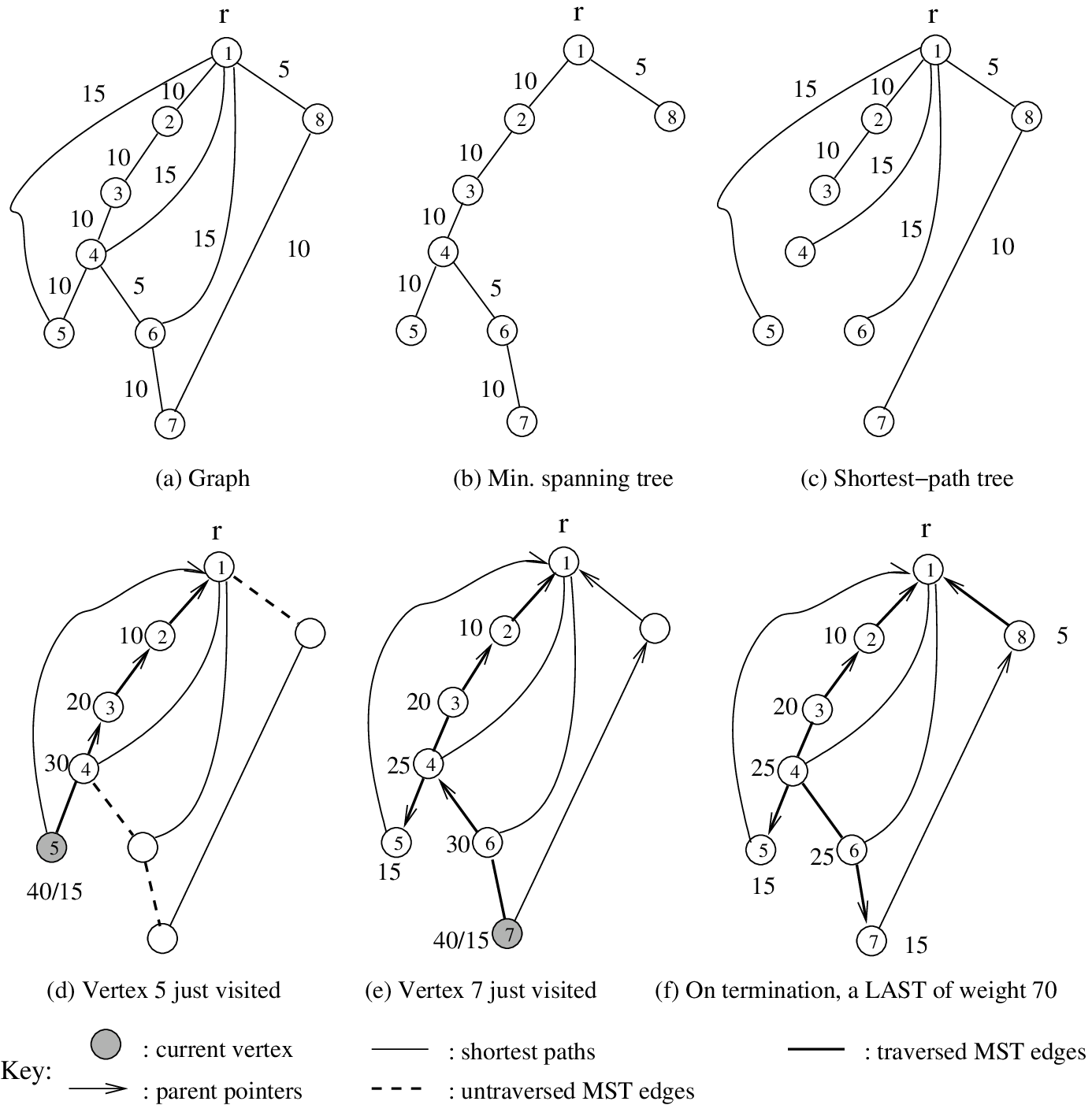}{figex}{A sample execution of the algorithm.}

Figure \ref{figex} shows a sample execution
of the algorithm with $\alpha = 2$
on the graph given in Frame (a).
Frames (b) and (c) give, respectively,
a minimum spanning tree (of weight 60)
and a shortest-path tree.

Initially all parent pointers are {\bf nil} and each $d[v]$ is infinite.
The depth-first search of the minimum spanning tree
visits the vertices in increasing order of their labels
and traverses the edges of the minimum-spanning tree
in the following order:
\[(1,2),(2,3),(3,4),(4,5),(5,4),(4,6),(6,7),
(7,6),(6,4),(4,3),(3,2),(2,1),(1,8),(8,1)\]
Recall that when an edge is traversed, it is relaxed.
When a vertex is visited, 
if its current distance estimate is not small enough
to guarantee the distance requirement then
the edges on the shortest path to the vertex are relaxed,
modifying the current tree.

Frame (d) shows the state of the algorithm
just after vertex 5 has been visited:
the edges $(1,2)$, $(2,3)$, $(3,4)$, and $(4,5)$ were relaxed
as they were traversed.
Because $d[5]$ was equal to 40 (more than twice the shortest-path distance)
when vertex $v$ was visited,
edge $(1,5)$ was relaxed, changing vertex $5$'s parent to vertex $1$, and
changing $d[5]$ to 15.

Frame (e) shows the state after vertex $7$ ---
the next vertex to have its shortest path added ---
has been visited.
Note that when edge $(5,4)$ was traversed, from $5$ to $4$,
its relaxation changed vertex $4$'s parent to vertex $5$
and updated $d[4]$ to reflect the new shorter path $(1,5),(5,4)$.
The algorithm then traversed and relaxed
edges $(4,6)$ and $(6,7)$,
bringing vertices 6 and 7 into the tree.
When vertex 7 was encountered, its distance
estimate (40) exceeded twice the shortest-path distance (15),
so the edges on the shortest path $(1,8),(8,7)$
to vertex $7$ were relaxed in that order.
This added these edges to the current tree
and brought down the distance estimates of these vertices.

Frame (f) shows the final state of the algorithm.
The parent pointers give the final tree.
Note that the relaxation of edge $(7,6)$  from 7 to 6,
changed vertex $6$'s parent.
This was the final change made to the tree.
Subsequent relaxations made by the traversal had no effect.
Remaining distance estimates were small enough to guarantee
that the distance requirements were met,
so that {\sc Add-Path} was not called.

\subsection{Analysis of the algorithm.  }
\label{analysisSec}
Next we prove that $\SC{Find-Last}(T_M, T_S, r, \alpha)$
returns an \OTL\ in linear time.
Let $T$ be the tree returned.

\begin{lemma} \label{distance_lemma}
The distance between $v$ and $r$ in $T$ 
is at most $\alpha$ times the shortest-path distance.
\end{lemma}
\begin{PROOF}
When a vertex $v$ is visited, 
if $d[v]$ exceeds $\alpha$ times the distance in the shortest-path tree then
{\sc Add-Path} is called, after which $d[v]$ equals the shortest-path distance.
In any case, 
after $v$ is visited, $d[v]$ is at most 
$\alpha$ times the shortest-path distance and subsequently never increases.
On termination it bounds the distance in $T$.
\end{PROOF}

An amortized analysis establishes 
that the total weight of $T$ is not too large.
\begin{lemma} \label{weight_lemma}
The weight of $T$ is at most
$\left(\minbeta\right)$ times the minimum spanning-tree weight.
\end{lemma}
\begin{PROOF}
Let $v_0 = r$ and let $v_1,v_2,...,v_k$ be the vertices 
that caused shortest paths to be added during the traversal,
in the order they were encountered.
When the shortest path from $r$ to $v_i$ $(i \ge 1)$ was added, 
the net weight of the added edges was at most $\dist{T_S}{r}{v_i}$.
Also, the edges on the path to $v_i$ 
consisting of the shortest path to $v_{i-1}$
followed by the path in the minimum spanning tree from $v_{i-1}$ to $v_i$
had been relaxed in order, so that
$d[v_i] \le \dist{T_S}{r}{v_{i-1}} + \dist{T_M}{v_{i-1}}{v_i}$.
The shortest path to $v_i$ was added because
$\alpha \dist{T_S}{r}{v_i} < d[v_i]$.
Combining the inequalities,
\[\alpha \dist{T_S}{r}{v_i} 
<  \dist{T_S}{r}{v_{i-1}} + \dist{T_M}{v_{i-1}}{v_i}.\]
Summing over $i$ bounds the net weight of the added paths:
\[
 \alpha\,\sum_{i=1}^{k} \dist{T_S}{r}{v_i}
 < \sum_{i=1}^{k} \left(\dist{T_S}{r}{v_{i-1}}+\dist{T_M}{v_{i-1}}{v_i}\right)
\]
and therefore
\[
 (\alpha-1)\,\sum_{i=1}^{k} \dist{T_S}{r}{v_i}
 < \sum_{i=1}^{k} \dist{T_M}{v_{i-1}}{v_i}.
\]
The DFS traversal traverses each edge exactly twice, 
and hence the sum on the right-hand side is at most
twice the weight of $T_M$, i.e.,
\[
 \sum_{i=1}^{k} \dist{T_M}{v_{i-1}}{v_i} 
 \le 2\,w(T_M).
\]
Hence the net weight of the added paths
is less than $\frac{2}{\alpha-1}w(T_M)$.
\end{PROOF}
The following alternate proof of Lemma \ref{weight_lemma}
may also be of interest.

{\em Alternate Proof of Lemma~\ref{weight_lemma}}.
As the algorithm executes,
define the potential function $\Phi$ 
to be the distance estimate of the current vertex.
When a shortest path of length $p$ 
to the current vertex $v$ is added,
$\Phi = d[v] > \alpha\, p$.
Adding the path lowers $d[v]$ to $p$,
decreasing $\Phi$ by at least $(\alpha - 1)\,p$.
Hence the total weight of the added paths
is bounded by the sum of the decrements
to $\Phi$ during the course of the algorithm,
divided by $\alpha - 1$.

Since $\Phi$ is initially 0 and always non-negative,
the sum of the decreases is at most the sum of the increments.
$\Phi$ increases only when the current vertex
changes from some vertex $u$ to a vertex $v$
after the edge $(u,v)$ was relaxed.
This ensures that $d[v] \le d[u] + w(u,v)$
and that $\Phi$ increases by at most $w(u,v)$.
Since each edge is traversed twice,
the total of the increases to $\Phi$ during the course
of the algorithm is bounded by 
twice the weight of the minimum spanning tree.

This establishes that the total weight of the added paths
is bounded by $\frac{2}{\alpha-1}$ times 
the weight of the minimum spanning tree.
\myendofproof

The running time is proportional to the number of relaxations.
This is $O(n)$  because each edge in $T_M$ or $T_S$
is relaxed at most twice by \SC{DFS}
and at most once by \SC{Add-Path}.
If the shortest-path tree and the minimum spanning tree
are not given, they can be computed in $O(m+ n\log n)$ time \cite{FT,GGST}.
This establishes Theorem \ref{mainthm}.

\noindent
{\bf Observation 1:}
In metric graphs
(complete graphs with edge weights satisfying the triangle inequality,
such as Euclidean graphs)
the shortest-path tree is trivial and can be found in $O(n)$ time.
For Euclidean graphs (induced by points in the plane)
the minimum spanning tree can  be computed 
in $O(n\log n)$ time \cite{PS}.  
In these cases the LAST can be found more quickly.
 
\noindent
{\bf Observation 2:} 
If the algorithm is given an {\em $a$-approximate} shortest-path tree
and a {\em $b$-approximate} minimum spanning tree,
the tree returned by the algorithm 
will be an $(a\alpha,b+2b/(\alpha - 1))$-LAST.
If such trees can be found more quickly
then a LAST can also be found more quickly.

\noindent
{\bf Observation 3:} 
In the {\em multiple-root} variant,
the distance requirement is that in the final tree (or forest)
the distance between each vertex and its {\em nearest} root
should be at most $\alpha$ times the distance to any root
in the original graph.
This variant can be easily reduced to the original problem
by adding an artificial root at distance 0 from the multiple roots.

\section{Optimality of the Algorithm}
\label{optimalitySec}
Next we show that the algorithm is optimal in the following sense.
Fix $\alpha>1$ and $1\le \beta<\minbeta$.
There is a planar graph not containing an {\ABL} 
rooted at a particular vertex.
Further, it is NP-complete to decide whether
a given graph contains an {\ABL}\ from a given root.

\fig{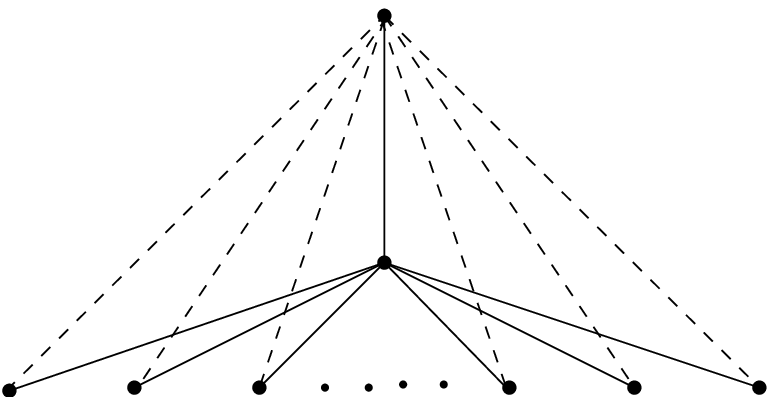}{fig4}{A graph with no \ABL\ for $\beta < \minbeta$,
($A=\alpha+1$, $B = \alpha + \epsilon - 1$ and $C = 2$).}

\subsection{Non-existence of LAST's when $\beta < \minbeta$.  }

\begin{lemma}
\label{notLeastThm}
If $\alpha > 1$ and $1 \leq \beta < \minbeta$,
then there exists a planar graph 
containing no \ABL\ rooted at a particular vertex.
\end{lemma}
\begin{PROOF}
The graph is shown in Figure {\ref{fig4}}. 
The structure of the graph is as follows.  
The root $r$ is connected to a central vertex $c$
by a path of weight $A$ of edges of weight some small $\delta$.
The central vertex is connected through similar paths 
of weight $B$ to the $\ell$ leaves.
The root is connected to each leaf with an edge of weight $C$.  
Let $A=\alpha+1$, $B = \alpha + \epsilon - 1$ and $C = 2$,
where $\epsilon$ is an arbitrarily small constant.  
For small enough $\delta$, the minimum spanning tree 
is formed by using all edges except those of weight~$C$.
Notice that this graph is planar.

Consider the paths from the root to any leaf.  
The shortest path is the direct edge of weight 2.
Any other path weighs more than $2 \alpha$:
the path through the center vertex weighs $A+B = 2 \alpha + \epsilon$;
any path through another leaf weighs at least
$2+2B = 2 (\alpha + \epsilon)$.
This means that in any \ABL\ all $\ell$ edges of weight 2 are present.
In addition, all but $\ell$ of the remaining edges are present.
Therefore the weight of any {\ABL}\ is at least $2\ell + T_M - \ell\delta$,
where $T_M = (\alpha+1) + \ell (\alpha- 1+ \epsilon)$ 
is the weight of the minimum spanning tree.
Hence the ratio of the weight of the \ABL\ to
the weight of the minimum spanning tree
is at least 
\[1+\frac{\ell(2-\delta)}{\alpha + 1 + \ell(\alpha - 1 + \epsilon)}\]
If $\beta < \minbeta$, then the above exceeds $\beta$ 
for sufficiently small $\epsilon$ and $\delta$ 
and sufficiently large $\ell$.  
\end{PROOF}

\subsection{NP-completeness of LAST queries.}
Next we show that
for any fixed $\alpha > 1$ and $1 \le \beta < \minbeta$
is  NP-hard to decide
whether a given graph contains an \ABL\ rooted at a given vertex.
Thus, it is unlikely that a polynomial-time algorithm
exists for finding \ABL's when $\beta < \minbeta$.

\fig{NPC-fig}{fig-npc}{Reduction From 3-SAT.}

Clearly the problem is in NP.
The proof of NP-hardness is in two parts. 
We first show NP-hardness for $\beta = 1$ and fixed $\alpha > 1$.
We then reduce this problem to the fixed $\beta < \minbeta$ case.

\begin{lemma}
\label{lem1}
For fixed $\alpha > 1$, 
deciding the existence of an $(\alpha, 1)$-LAST 
rooted at a given vertex of a given graph is NP-hard.
\end{lemma}
\begin{PROOF}
The proof is by reduction from 3-SAT.
Let $F$ be a $3$-SAT formula in conjunctive normal form ---
each clause consists of three literals from 
$\{x_1, \ldots, x_n \} \cup \{ \overline{x}_1, \ldots, \overline{x}_n \}$.
We build a graph in which the 
$(\alpha,1)$-LAST's correspond to satisfying assignments of $F$.

$A$, $B$, $D$, $E$ and $W$ are constants to be determined later.
The graph has a root vertex $R$, a vertex $S$,
and a path connecting $R$ to $S$ of weight $D$
consisting of edges small enough to ensure that
the path is in any minimum spanning tree.

For each pair of literals $x_i$ and $\overline{x}_i$, 
there are two vertices $X_i$ and $\overline{X}_i$,
each having an edge to $S$ of weight $A$.
A path of weight $E$ connects $X_i$ and $\overline{X_i}$.
This path is also constructed so as to be in any minimum spanning tree.

For each clause $c_j$ there is a vertex $C_j$
with an edge to $R$ of weight $W$.
From $C_j$ to each vertex corresponding to a literal in $c_j$
there is an edge of weight $B$.

This defines the graph.  
Observe that, provided $0 < A < B < W$, 
the minimum spanning trees are exactly characterized by the following.
In any minimum spanning tree, 
the path from $R$ to $S$ 
and each path from $X_i$ to $\overline{X}_i$ are present.
For each variable $x_i$,
exactly one of the two edges $\{ (S,X_i), (S,\overline{X}_i) \}$ is present.
For each clause $c_j$,
exactly one edge of the form $(X_i,C_j)$ or $(\overline{X}_i,C_j)$
for some $i$ is present.
No other edges are present.

Next we use the distance requirement
to ensure that any minimum spanning tree is an $(\alpha,1)$-LAST
if and only if the edge to each clause vertex 
comes from some variable vertex $X_i$ or $\overline{X}_i$
that has an edge in the minimum spanning tree directly to $S$.
This is all that is needed,
for then the $(\alpha,1)$-LAST's will correspond 
to satisfying assignments in the original formula,
and vice versa, as follows: 
for each variable $x_i$, choose the edge $(S,X_i)$
iff $x_i$ is true, otherwise choose the edge $(S,\overline{X}_i)$;
for each clause $c_j$, choose the edge $(X_i,C_j)$
(or $(\overline{X}_i, C_j)$), 
where $x_i$ (or $\overline{x}_i$) is a variable
(or negated variable) satisfying $c_j$.

It suffices to choose $A$, $B$, $D$, $E$ and $W$ so that
\[ 0 < A < B < W \]
\[ D + A + E \leq \alpha \min\{A+D, B+W\} \]
\[ D + A + B \leq \alpha \min\{D+A+B, W\} < D+A+E+B. \]
To achieve this, let
$A = 1$,
$B = \alpha$,
$D = 2 \alpha$,
$E = (\alpha -1)(2 \alpha +1)$ and
$W = 1 + 2 \alpha + \frac{1}{\alpha}$.
\end{PROOF}

Next we will reduce the $(\alpha,1)$-LAST problem
to the $(\alpha, \beta)$-LAST problem,
for any fixed $\alpha$ and $\beta$
such that $\alpha > 1$ and $1 \leq \beta < \minbeta$.

\begin{PROOF}(Theorem \ref{optimalityThm})
Let $G^*$ be the graph 
for which we want to determine the existence
of an $(\alpha,1)$-LAST rooted at a given vertex $r^*$.
By Lemma~\ref{notLeastThm}, 
there exists a graph $G^\prime$ with no {\ABL}\ rooted
at some vertex $r'$.
Assume without loss of generality that 
the minimum spanning tree of $G^*$ has weight $1$
and the minimum spanning tree of $G^\prime$ 
is of weight $c$ (a constant to be determined later).
Define the graph $G$ to be the union of $G^*$ and $G^\prime$
by identifying $r^*$ and $r'$ into a single root $r$.

Let $\beta^\prime$ be the minimum $\beta$
such that $G^\prime$ has an \ABL.
Define $\beta^*$ analogously for $G^*$. 
Take $c=\frac{\beta-1}{\beta^\prime-\beta}$.

The weight of the minimum spanning tree in $G$ is $1+c$;
similarly the lightest tree in $G$
meeting the distance requirement
is of weight $\beta^* + \beta^\prime\,c$.
Thus $G$ has an \ABL\ iff $\beta^* + \beta^\prime\,c \le \beta(1+c)$.
By our choice of $c$, this is equivalent to
$\beta^* \le 1$.
Thus $G$ has an \ABL\ iff $G^*$ has an $(\alpha,1)$-LAST.
\end{PROOF}

\section{Minimum-Weight Shortest-Path Trees}
\label{minWtSPTSec}
Next we consider the case when $\alpha=1$, i.e.,
an {\ABL}\ is a shortest-path tree
of weight at most $\beta$ times the weight of the minimum spanning tree.
In this case, no algorithm can guarantee any fixed $\beta$ for all graphs.  
Instead, we show how to find a $(1,\beta)$-LAST
with minimum $\beta$ in a given graph, i.e.,
a minimum-weight shortest-path tree.

In fact, we solve a more general problem:
finding a minimum-weight shortest-path tree
in a rooted {\em directed} graph.
The undirected case reduces to this case
by the standard trick of replacing each undirected edge $(u,v)$
by two new directed edges $(u,v)$ and $(v,u)$ of the same weight
as the original edge.

The directed problem reduces in turn
to the problem of finding a {\em minimum-weight branching}
in the {\em shortest-path subgraph} of the given directed graph.
A branching is a directed spanning tree with all edges
directed away from the root.
The shortest-path subgraph
is the spanning subgraph consisting of all directed edges $(u,v)$ 
on some shortest path from the root,
i.e., those for which $\dist{G}{r}{u} + w(u,v) = \dist{G}{r}{v}$.
It is easy to show that the shortest-path trees 
in a directed graph are exactly the branchings from the root
in its shortest-path subgraph.
Consequently, it suffices to find a
minimum-weight branching in the shortest-path subgraph.

A polynomial-time algorithm for finding a minimum-weight branching
in {\em any} given graph is known \cite{GGST}.
However, a shortest-path subgraph of a non-negatively weighted graph
has the property that any edge on a cycle has weight zero.
This allows the following linear-time algorithm.
First, identify the strongly-connected components 
in the subgraph induced by the edges of weight zero.
This can be done in linear time \cite{CLR}.
For each component not containing the root,
choose the minimum-weight incoming edge
and call the vertex with an incoming chosen edge
the {\em base} vertex of the component.
For the component containing the root,
call the root vertex the base vertex.
For each component, find a branching
of weight zero edges rooted at the base
in the subgraph induced by the component.
Finally, return the chosen edges
together with the edges of the components' branchings.

This set of edges forms a branching:
each non-root vertex has an incoming edge
and there are no cycles.
The branching is of minimum weight
because in {\em any} branching every non-root component
has at least one incoming edge.
It is straightforward to implement the algorithm to run in $O(n+m)$ time.
This proves Theorem \ref{minWtSPTThm}.

\section{Finding LAST's in Parallel}
\label{parallelSec}
Given $\alpha>1$, a minimum spanning tree, and a shortest-path tree, 
an \OTL\ can be found using $n$ processors in $O(\log n)$-time.
The model of computation we use is the {\em Concurrent-Read,
Exclusive-Write Parallel RAM}, in which independent,
synchronized parallel processors share a common memory  \cite{JaJa}.
Multiple simultaneous accesses to the same memory location
are allowed only if all of the accesses are read operations.

The algorithm is as follows.
Let $C=(e_1,e_2,...,e_{2n-2})$ be the (directed) edges 
of the walk through the graph implicit in the depth-first search
of the minimum spanning tree, as in Section \ref{relaxSec}.
This tour can be constructed in $O(\log n)$ time 
by $n$ processors using standard techniques~\cite{JaJa}. 
Let $(u_i,u_{i+1})=e_i$.
Using the terminology of Section \ref{relaxSec},
after edge $(u_i,u_{i+1})$ is traversed from $u_i$ to $u_{i+1}$,
vertex $u_{i+1}$ is the current vertex.

The parallel algorithm emulates the serial algorithm
except that the distance estimates are looser in two ways.
First, while a vertex may occur several times in $C$,
the parallel algorithm treats each occurrence as a distinct vertex.
Second, when a shortest path is added, 
only the distance estimate of the destination vertex is lowered.
These looser distance estimates can be computed in parallel,
but still suffice to imply the weight requirement.

Let $\dist{C}{u_i}{u_j}$ denote $\sum_{k=i}^{j-1} w(e_k)$, 
the distance from $u_i$ to $u_j$ along $C$. 
Let $m(i,j)$ be the relation
\[i < j\ \ \mbox{and}\ \ 
\dist{T_S}{r}{u_i}+\dist{C}{u_i}{u_j} > \alpha\,\dist{T_S}{r}{u_j}.\]
The meaning of $m(i,j)$ is the following.
Suppose we modify the original algorithm 
to use the looser distance estimates.
When the modified algorithm enountered a vertex $u_j$,
if the algorithm had not added shortest paths
to any of the vertices $u_{i+1}, u_{i+2}, ..., u_{j-1}$,
then it would add the shortest path from the root to $u_j$.
Thus, if the modified algorithm adds a shortest path
to a vertex $u_i$, then the next shortest path it adds
will be to vertex $u_k$, where $k=\min\{j : m(i,j)\}$.

The parallel algorithm will emulate the modified algorithm.
Define $J(i) = \min\{ j : m(i,j)\}$. 
The parallel algorithm will compute the function $J$
and then add shortest paths to vertices in the set
\[S=\{u_1,u_{J(1)},u_{J(J(1))}, ..., u_{J(J(...J(1)...))}\}.\]

Once $J$ has been computed, 
$S$ can be computed by $n$ processors in $O(\log n)$ times
on a CREW PRAM 
using standard techniques \cite{JaJa}.
Once $S$ has been computed,
the set $S^*$ of ancestors of $S$ in the shortest-path tree
can also be computed
by $n$ processors in $O(\log n)$ time 
using tree contraction techniques \cite{JaJa}.
The final tree is formed by each non-root vertex choosing as its parent
either the parent in the shortest-path tree
(if the vertex is in $S^*$)
or the parent in the minimum spanning tree
(otherwise).
It can easily be shown that every vertex has a path to the root
using this set of $n-1$ edges, so that they do indeed form a tree.

It remains to compute $J(i)$.
First, note that $m(i,j)$ is monotone in $i$ for fixed $j$.
\begin{lemma} \label{monotonicity-lemma}
If $i' < i$ and $m(i,j)$ is true then $m(i^\prime,j)$ is true.
\end{lemma}
\begin{PROOF}
If $m(i,j)$ is true then
$\dist{T_S}{r}{u_i}+\dist{C}{u_i}{u_j} > \alpha\,\dist{T_S}{r}{u_j}$. 
For any $i^\prime<i$, $\dist{C}{u_i}{u_j} \le \dist{C}{u_i^\prime}{u_j}$.
The shortest path from $r$ to $i$ is no longer than any other
path from $i$ to $r$ in the graph, and hence
$\dist{T_S}{r}{u_i} \le \dist{T_S}{r}{u_{i^\prime}} + 
\dist{C}{u_i^\prime}{u_i}$.
Combining these inequalities, we get
\begin{eqnarray*}
\dist{T_S}{r}{u_{i^\prime}} + \dist{C}{u_i^\prime}{u_j} 
        &=& \dist{T_S}{r}{u_{i^\prime}}+\dist{C}{u_i^\prime}{u_i}+
\dist{C}{u_i}{u_j} \\
        &\ge& \dist{T_S}{r}{u_i} + \dist{C}{u_i}{u_j}\\
        &>&   \alpha\,\dist{T_S}{r}{u_j}
\end{eqnarray*}
Hence $m(i^\prime,j)$ is true by definition.
\end{PROOF}

The function $J$ can be computed efficiently
because of this monotonicity property:
\begin{lemma}
Suppose $m(i,j)$ implies $m(i',j)$ 
for $0 \le i' \le i \le n$, $0 \le j \le n$.
Then the function $J(i) = \min\{j : m(i,j)\}$
can be computed in $O(\log n)$ time
by $n$ processors on a CREW PRAM.
\end{lemma}
\begin{PROOF}
Define $I(j) = \max\{i : m(i,j)\}$.
For each $j$, compute $I(j)$ using binary search.
Define $I^*(j) = \max\{I(j') : 1 \le j' \le j\}$.
Compute function $I^*$ from function $I$ using 
a standard prefix-maxima computation.
Finally, define $J'(i) = \min\{j : I^*(j) \ge i\}$.
Compute function $J'$, again using binary search,
from monotone function $I^*$.
(See Figure \ref{parFig}.)
Each of these computations can be done
by $n$ processors in $O(\log n)$ time
on a CREW PRAM using standard techniques \cite{JaJa}.

\fig{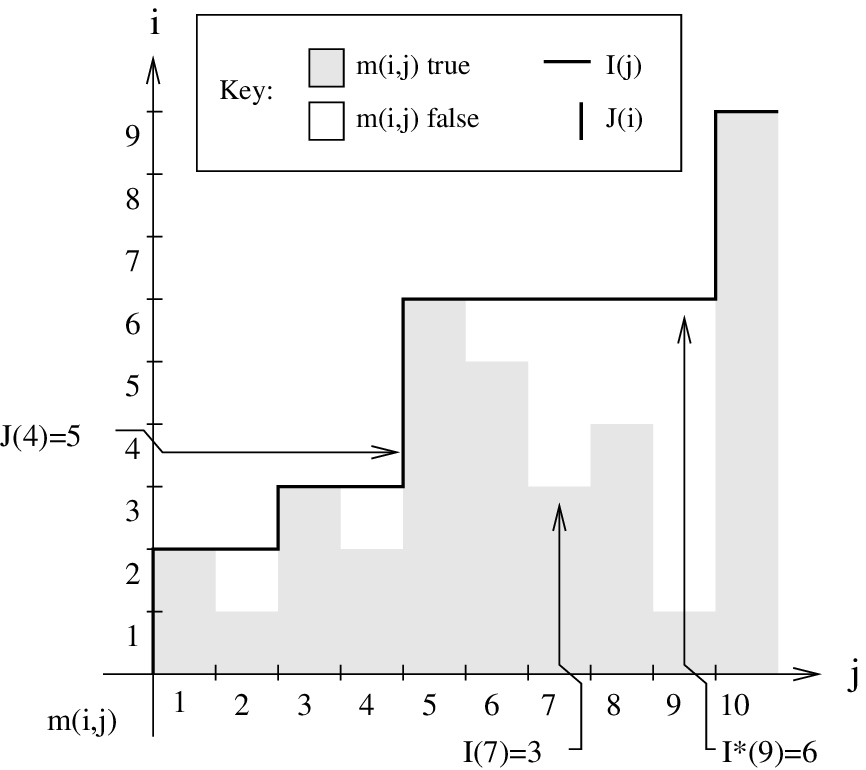}{parFig}{Computing $J$ from $m$ to find a LAST in parallel.}
We prove that $J'(i) = J(i)$ for each $i$.
The proof is in two steps.
\begin{enumerate}
\item $J'(i) = \min\{j : I^*(j) \ge i\} = \min\{j : I(j) \ge i\}$,
i.e., the smallest $j$ such that $I^*(j)$ exceeds $i$ is
equal to the smallest $j$ such that $I(j)$ exceeds $i$.
This is because the latter depends only on the maxima of $I$ ---
those $j$ such that $I(j) > I(j')$ for all $j' \le j$.
\item $\min\{j : I(j) \ge i\} = \min\{j : m(i,j)\}$
because $I(j) \ge i$ is equivalent to $m(i,j)$
by the monotonicity property of $m$.
\end{enumerate}
\end{PROOF}

The analyses of Lemmas \ref{distance_lemma} and \ref{weight_lemma}
can easily be adapted 
to prove that the final tree produced by the parallel algorithm is an \OTL.
This establishes Theorem \ref{parallelThm} ---
an \OTL\ can be computed 
by $n$ processors in $O(\log n)$ time on a CREW PRAM.

\section{Conclusions}
Every graph contains trees 
that offer a continuous tradeoff 
between minimum spanning trees and shortest-path trees. 
Trees achieving the optimal trade-off
can be found in (sequential) linear time
or in logarithmic time by linearly many processors.

In the following cases, is it possible to obtain a better trade-off?
\begin{itemize}
\item In Euclidean graphs (i.e., points in the plane).
Note that the proof of Lemma \ref{weight_lemma}
requires only that the algorithm walk around
the graph from the root visiting every vertex once,
i.e., that the algorithm traverse 
a Traveling Salesman path starting at the root.
In Euclidean graphs, perhaps such a path
of weight at most $(2-\epsilon)$ 
times the minimum spanning-tree weight always exists
and can be found in polynomial time.

\item If the distance requirement is replaced
by the requirement that the {\em sum} of distances from the root
is within $\alpha$ times the minimum possible?

\item If the root is not fixed?
This would correspond to the problem
of installing a low-cost network {\em and} choosing a root site
so that distances from the root are near-minimum.
\end{itemize}
Clearly any \ABL\ also meets these looser requirements,
but our lower bounds no longer show that the trade-off is optimal.

For directed graphs, it is easy to show that
for any fixed $\alpha$ and $\beta$, \ABL's may not exist
and that finding the minimum $\beta$ such that an \ABL\ exists is NP-hard.
Can one {\em approximate} this minimum $\beta$?
 
\section*{Acknowledgments}
We would also like to thank Seffi Naor, Dheeraj Sanghi and Moti Yung for 
useful discussions.
We would like to thank Shay Kutten for telling us about {\cite{ABP1}}.
We would like to thank Baruch Awerbuch, Alan Baratz and David Peleg
for sending us a copy of their manuscript \cite{ABP2}.
We would like to thank Andrew Kahng and Jeff Salowe for telling us about 
{\cite{CKRSW1,CKRSW2,CKRSW3}}.

\end{document}